# The Efficiency Costs of Information Assurance in AI-Enabled Labor Markets: Evidence from LinkedIn's Policy Changes

**Lei Chen**

School of Management, University of Science and Technology of China

leichen1109@mail.ustc.edu.cn

**Chaoyue Gao**

School of Management, University of Science and Technology of China

gaochaoyue@ustc.edu.cn

**Alvin Leung**

College of Business, City University of Hong Kong

acmleung@cityu.edu.hk

**Gavin Wang**

Mitch Daniels School of Business, Purdue University

Xiaoning.Wang@purdue.edu

# The Efficiency Costs of Information Assurance in AI-Enabled Labor Markets: Evidence from LinkedIn's Policy Changes

**Abstract**

Generative artificial intelligence (GenAI) systems rely heavily on user-generated data for training. As governments and platforms impose increasing restrictions on the use of personal data, an important question is whether limiting access to user data for AI training affects the performance of AI-enabled economic systems. We examine this question in the context of labor-market matching. Our setting exploits a unique sequence of LinkedIn policy changes: the quiet introduction of user data collection for AI training in August 2024, the restriction of Hong Kong user data from AI training in October 2024, and the subsequent restoration of data access in November 2025. Using employment and job-posting data from Revelio Labs and a Difference-in-Differences design comparing Hong Kong and Singapore, we find that the restriction significantly increased labor-market frictions: employee turnover increased and tenure declined, vacancies remained open longer, job-posting match rates fell, and wages decreased. To strengthen the causal interpretation, we examine the full policy lifecycle: labor market outcomes improved following LinkedIn's initial AI rollout, deteriorated after the restriction, and recovered after the restoration. Similar patterns emerge in an independent cross-country comparison between regions with and without access to LinkedIn's user data for AI training. The effects are strongest among firms that rely more heavily on platform-mediated recruitment. Our findings show that restricting access to user data for AI training can affect labor-market outcomes through AI-enabled matching systems and contribute to ongoing discussions regarding AI governance, privacy regulation, and the economic consequences of data access.

# 1. INTRODUCTION

Generative artificial intelligence (GenAI) is increasingly embedded in digital platforms that mediate economic activity. Because these systems rely on large volumes of user-generated data for training, policymakers and platform operators face a fundamental challenge: restrictions on data access may enhance privacy protection but may also reduce the effectiveness of AI-enabled services (Acquisti et al., 2015; Acquisti et al., 2016; Ghose et al., 2026; Godinho de Matos & Adjerid, 2022; Sun et al., 2024). As privacy and data-governance rules tighten worldwide, understanding the economic consequences of limiting user data for AI training has become an important managerial and policy question.

The labor market represents a particularly high-stakes context for this trade-off. Platforms such as LinkedIn increasingly use AI to support candidate discovery, job recommendations, recruiter assistance, and job–candidate matching (Faberman & Kudlyak, 2016; Kokkodis & Ipeirotis, 2023). Prior research on pre-GenAI online search technologies, however, finds limited or mixed effects on aggregate labor-market outcomes, with online tools often substituting for rather than expanding traditional search channels (Kroft & Pope, 2014; Kuhn & Mansour, 2014; Kuhn & Skuterud, 2004). Unlike earlier online tools that merely lowered search costs, GenAI-powered matching actively participates in candidate–job pairing decisions. This qualitative shift enhances matching efficiency, but it also creates a critical vulnerability: its decision-making capabilities are directly dependent on large-scale, continuously updated user data. As a result, restrictions on training data are likely to be far more consequential for GenAI than they were for previous technologies. Recent research suggests that limiting access to fresh human-generated data can reduce the quality of AI outputs and impair model performance (e.g., Shumailov et al., 2024; Lu et al., 2023). Privacy-preserving techniques designed to protect user data can themselves impose measurable costs on model utility, illustrating a broader trade-off between data protection and AI performance (Zhang et al., 2025). Yet despite growing interest in AI governance and privacy regulation, there is little empirical evidence on how restrictions on AI-training data affect labor-market outcomes in real-world settings, making it difficult for policymakers to weigh the economic costs against privacy gains in this high-stakes domain.

We study this question using a unique policy sequence on LinkedIn. In August 2024, LinkedIn quietly began using member data to train its generative AI models. Following concerns raised by privacy regulators, the platform suspended the use of Hong Kong users' data for AI training on October 11, 2024, while no analogous restriction was imposed in Singapore. Subsequently, on November 3, 2025, LinkedIn restored user data access for AI training in Hong Kong as part of its updated global data-use policy. This introduction–restriction–restoration sequence provides a rare opportunity to examine how changes in AI-training-data availability affect labor-market matching outcomes.

Using large-scale employment and job-posting data from Revelio Labs, we compare firms in Hong Kong and Singapore using a Difference-in-Differences design. We find that the restriction significantly increased labor-market frictions. Employee turnover increased and tenure declined, vacancies remained active for longer periods, job-posting match rates fell, and wages decreased. We then provide a series of validation tests linking these outcomes to user data access for AI training. Matching outcomes improve following LinkedIn's initial AI rollout, deteriorate after the Hong Kong restriction, and subsequently recover after the restoration. Similar patterns also emerge in an independent comparison between countries where LinkedIn retained user data access for AI training and countries where such access remained restricted.

This study contributes to the growing literature on AI governance, privacy regulation, and digital platforms (Godinho de Matos & Adjerid, 2022; Godinho de Matos et al., 2018; Zhang et al., 2021) by providing empirical evidence on the labor-market consequences of restricting access to user data for AI training. Leveraging a rare policy sequence involving the introduction, restriction, and restoration of data access, we quantify how changes in training-data availability affect worker mobility, employment duration, vacancy filling, and wages. Because labor markets constitute a high-stakes economic infrastructure, even modest changes in matching efficiency can have meaningful implications for firms, workers, and aggregate productivity. Our findings therefore provide empirical evidence relevant to ongoing discussions regarding the economic tradeoffs of privacy regulation and data governance in AI-enabled platforms.

## 2. RESEARCH CONTEXT AND DATA

### 2.1 The Three-Stage LinkedIn AI Policy Timeline

A central challenge in studying the labor market impact of AI-driven matching is identifying exogenous variation in the availability of training data. Because AI systems typically evolve continuously and their training processes are proprietary, it is often difficult to isolate when and how changes in data access affect matching performance. The recent AI policy changes of LinkedIn provide a rare opportunity to do so. Beginning in 2024, LinkedIn introduced a series of policy changes governing whether members' personal data could be used to train the platform's GenAI models in some regions. These GenAI models are central to LinkedIn's job-matching ecosystem: they power job recommendations for candidates and candidate suggestions for recruiters, directly shaping the matching process between firms and workers.[1] Consequently, the quality of LinkedIn's GenAI models directly affects job matching quality and overall platform performance. These policy changes created three distinct phases: introduction, restriction, and restoration of user data access for AI training, allowing us to observe how matching outcomes responded as the platform's access to training data expanded, contracted, and then recovered. Figure 1 summarizes this policy timeline.

[1] Sources: https://www.linkedin.com/blog/engineering/infrastructure/incremental-and-online-training-platform-at-linkedin, https://www.linkedin.com/blog/engineering/ai/jude-llm-based-representation-learning-for-linkedin-job-recommendations and https://www.linkedin.com/blog/engineering/ai/building-the-next-generation-of-job-search-at-linkedin.

Figure 1. The Three-Stage LinkedIn AI Policy Timeline in Hong Kong

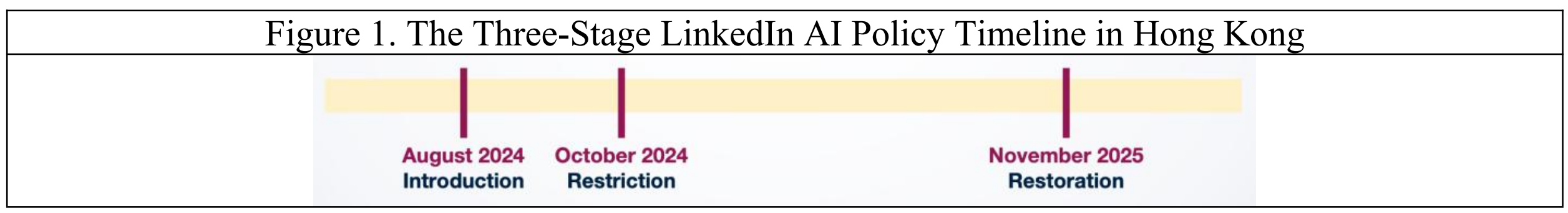

**Introduction:** In August 2024, LinkedIn quietly updated its privacy settings to allow members' personal data, including posts, activity histories, and profile information, to be used for training its GenAI models. The setting was enabled by default, automatically enrolling users without requiring affirmative opt-in consent.[2]

**Restriction:** In October 2024, amid growing regulatory scrutiny and public concern regarding the use of personal data for AI training, LinkedIn suspended AI-training data collection in several jurisdictions. In Hong Kong, the Office of the Privacy Commissioner for Personal Data (PCPD) formally raised concerns regarding LinkedIn's use of personal data for AI model development. Following these discussions, LinkedIn ceased using Hong Kong members' personal data for AI training effective October 11, 2024.[3]

**Restoration:** Between October 2024 and April 2025, Hong Kong regulators engaged with LinkedIn to review its data-use proposal, and LinkedIn progressively reached privacy agreements with various jurisdictions. On November 3, 2025, LinkedIn resumed using member data in Hong Kong to train its GenAI models.[4]

This three-stage policy sequence is central to our identification strategy. Most studies of AI policy rely on a single treatment event and therefore face concerns that observed changes may reflect coincident trends or unrelated shocks. In contrast, LinkedIn's policy generated a sequence of symmetric changes in data availability: access to training data was first expanded, then restricted, and later restored. This allows us to examine whether matching outcomes improve following the introduction of AI-training data, deteriorate following the restriction, and recover after restoration. Observing this predicted pattern across all three stages would be difficult to reconcile with alternative explanations unrelated to the platform's AI training data policy.

---

[2] The policy was quietly implemented uniformly across all countries and regions where LinkedIn operates, excluding the European Union, the European Economic Area (comprising Iceland, Liechtenstein, and Norway), and Switzerland due to compliance with the General Data Protection Regulation (GDPR). Sources: https://www.reuters.com/legal/microsofts-linkedin-sued-disclosing-customer-information-train-ai-models-2025-01-22

[3] Source: https://www.pcpd.org.hk/english/news_events/media_statements/press_20241015.html.

[4] Source: https://www.pcpd.org.hk/english/news_events/media_statements/press_20251030.html.

### 2.2 Hong Kong and Singapore: Treatment and Control Markets

To identify the impact of LinkedIn's Hong Kong-specific data restrictions and their subsequent restoration, we employ a difference-in-differences (DiD) design that treats Hong Kong as the affected market and Singapore as the control market. Singapore provides a natural benchmark because it closely resembles Hong Kong in terms of economic structure, labor-market composition, and LinkedIn user characteristics. Both are highly developed, service-oriented international business hubs with highly educated, English-speaking workforces. As shown in Figure 2, LinkedIn user demographics are remarkably similar across the two markets, particularly in terms of age composition.

Figure 2. LinkedIn Users Composition: Hong Kong vs. Singapore[5]

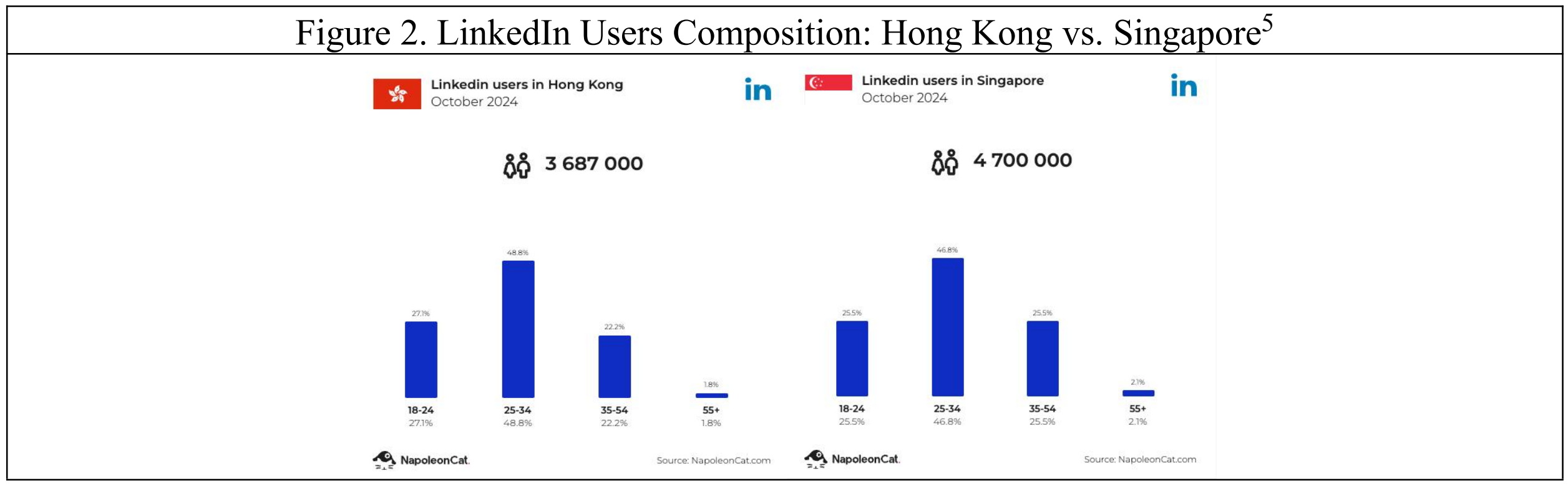

As a further robustness check, we leverage a broader cross-country comparison by contrasting countries where LinkedIn retained access to user data for AI training (e.g., the United

[5] Sources: https://napoleoncat.com/stats/linkedin-users-in-hong_kong/2024/10/ and https://napoleoncat.com/stats/linkedin-users-in-singapore/2024/10/.

States and the United Kingdom) with European Union countries, where such access remained restricted under GDPR. Details are discussed in Section 5.2.

### 2.3 Labor Market Data: Revelio Labs

We measure labor-market outcomes using data from Revelio Labs, which integrates professional profile histories and firm-level job postings from multiple online sources. According to the company, the dataset covers approximately 76% of the labor force in Hong Kong and 83% in Singapore.[6] We verified that coverage remains stable throughout our sample period, with no discontinuity around the policy intervention in either market. Importantly, LinkedIn's restriction on the use of personal data for AI training did not affect Revelio's data collection process, as Revelio independently aggregates information from publicly available professional profiles. As a result, the observed changes in labor market outcomes are unlikely to be driven by shifts in Revelio's underlying data coverage or measurement process. Our analysis draws on three Revelio Labs datasets. The *Workforce Dynamics* table provides firm-month measures of employment duration and worker inflows and outflows. The *Job Postings* table contains posting and removal dates as well as job characteristics. The *Individual Positions* table records worker-level employment histories, which we link to job postings using firm identifiers, posting timing, and standardized job classifications to construct matching outcomes.

Our sample spans 30 months, covering the 12 months preceding the October 2024 restriction and the subsequent 18 months. This observation window captures the full sequence of LinkedIn's AI policy changes, including the August 2024 introduction of user data collection for AI training, the October 2024 restriction in Hong Kong, and the November 2025 restoration of data access. As a result, we are able to observe labor-market outcomes from the initial rollout of AI training through five months after the restoration, allowing us to examine the complete introduction–restriction–restoration cycle.

---

[6] Source: Revelio Profiles Coverage by Country: https://wrds-www.wharton.upenn.edu/pages/support/manuals-and-overviews/revelio-labs/.

## 3. VARIABLES AND EMPIRICAL STRATEGY

### 3.1 Variable Definitions

In this study, we distinguish between matching quality and matching efficiency: Matching quality captures the durability and stability of employer–employee matches after hiring, whereas matching efficiency captures how quickly vacancies are resolved and whether postings successfully result in hires. Accordingly, we use Flux Rate and Average Duration as measures of matching quality, and Active Posting Period and Matched Job Postings Rates as measures of matching efficiency.

**Flux Rate and Related Turnover Measures:** We construct **Inflow Rate**, **Outflow Rate**, and **Flux Rate** as employee movements scaled by firm employment. Specifically, Inflow Rate is the number of employees joining from external organizations divided by firm employment, Outflow Rate is the number leaving for external organizations divided by firm employment, and Flux Rate is the sum of the two. Expressing worker movements as rates facilitates comparisons across firms of different sizes (Davis & Haltiwanger, 1999; Kali & Liu, 2024; Li et al., 2022).

**Average Duration:** Revelio Labs constructs employment spell durations using a proprietary machine-learning model that infers the start and end dates of positions from professional profile histories. We aggregate these durations to the firm-month level to construct **Average Duration**, which measures the average realized tenure of employees associated with a firm. A decline in this measure indicates that departing employees leave after shorter employment spells on average. Because successful employment matches tend to persist longer, while mismatches are more likely to terminate early (Jovanovic, 1979; Pissarides, 2011), shorter realized tenure is a signal of less durable employer–employee matching. In our context, a reduction in Average Duration is therefore consistent with a decline in matching quality following the restriction.

**Active Posting Period: Active Posting Period** measures the number of days between a job posting's publication and removal (Davis et al., 2012, 2013). Revelio does not distinguish whether a posting was removed because the position was filled, withdrawn, or expired; all are recorded as closed postings. Accordingly, this measure should be interpreted as the time a vacancy remains actively advertised on the platform rather than a direct measure of successful hiring. Although the underlying reason for removal is unobserved, shorter active posting periods are generally consistent with improved matching efficiency.

**Matched Job Postings Rate:** To construct **Matched Job Postings Rate**, we link job postings to newly observed employment positions within the same firm using posting timing and standardized job-title and job-description classifications (Marinescu & Wolthoff, 2020). A posting is considered matched if at least one qualifying employee record appears after the posting date. We also construct **Half-Year Matched Job Postings Rate**, which requires the match to occur within six months of posting.

**Moderators: Firm Size, Seniority, and Platform Dependence.** We examine three moderators that may shape firms' dependence on platform-based matching. **Firm Size** is measured as the number of employees in a firm-month and captures organizational scale. Larger firms typically possess stronger recruiting resources, employer-brand recognition, and alternative hiring channels, potentially reducing their reliance on platform algorithms. **Seniority** measures the average hierarchical rank of employees within a firm-job-month cell based on Revelio's seven-level job hierarchy classification. Firms with more senior workforces may similarly be less dependent on platform-mediated matching, as experienced workers are more likely to rely on professional networks and direct recruitment channels rather than online job platforms (Garg & Telang, 2018). **Dependence** measures the proportion of employees in a firm-month hired from external recruitment platforms in the three months before the restriction. Firms with higher dependence on external recruitment platforms are more likely to be impacted by the platform policy changes.

To control for potential time-varying market conditions, we also control for monthly regional GDP, Rental Price Index, Population, Unemployment Rate, and CPI Growth Rate. Table 1 reports the main variable definitions used in the analysis and their descriptive statistics, based on the sample for the main analysis (12 months before and after the restriction date).

Table 1. Main Variable Definitions

| Variable type | Variable name | Variable definition | Mean | Std. Dev. |
|---|---|---|---|---|
| Dependent Variables (Matching Quality) | Flux Rate$_{i,t}$ | The sum of employee inflows from outside organizations and outflows to outside organizations for firm $i$ in month $t$, divided by the number of employees. | 0.0136 | 0.0604 |
| | Average Duration$_{i,t}$ | Average duration (days) of all positions at firm $i$ until month $t$ | 572.7587 | 871.5747 |
| Dependent Variables (Matching Efficiency) | Active Posting Period$_{i,t}$ | Average number of days from posting to removal for positions posted by firm $i$ in month $t$. | 31.0363 | 35.0843 |
| | Matched Job Postings Rate$_{i,t}$ | Rate of positions posted by firm $i$ in month $t$ that were actually filled by new hires. | 0.2494 | 0.3760 |
| | Half-Year Matched Job Postings Rate$_{i,t}$ | Rate of positions posted by firm $i$ in month $t$ that were actually filled by new hires within 6 months | 0.2069 | 0.3454 |
| Independent variables | Treat$_i$ | Firm location dummy variable (1 = Hong Kong, 0 = Singapore) | 0.4713 | 0.4992 |
| | Post$_t$ | Policy intervention period dummy variable | 0.5000 | 0.5000 |
| Moderator variables | Size$_{i,t}$ | Number of employees at firm $i$ in month $t$ | 19.0910 | 251.4833 |
| | Seniority$_{i,j,t}$ | Average hierarchical rank of employees at firm $i$, job $j$, month $t$ | 2.7470 | 1.2610 |
| | Dependence$_{i,t}$ | Percentage of employees hired from external recruitment platforms in the three months before the restriction | 0.7569 | 5.8856 |

### 3.2 Empirical Strategy and Identification Challenges

Our unit of analysis is the firm-month level. To estimate the effect of LinkedIn's restriction on the use of Hong Kong members' data for AI training, we employ a Difference-in-Differences (DiD) design comparing firms located in Hong Kong with firms located in Singapore. Because labor-market outcomes exhibit strong seasonal patterns, we use a symmetric estimation window consisting of the 12 months before and the 12 months after the October 11, 2024 restriction. Our baseline specification is:

$$Y_{i,t} = \beta_0 + \beta_1 ( Treat_i \times Post_t ) + u_i + \tau_t + \epsilon_{i,t} \qquad (1)$$

where $Y_{i,t}$ denotes one of the matching outcomes defined in Table 1. $Treat_i$ equals one for firms located in Hong Kong and zero otherwise, while $Post_t$ equals one for observations after the restriction and zero beforehand. Firm fixed effects ($u_i$) absorb time-invariant differences across firms, and month fixed effects ($\tau_t$) absorb common labor-market shocks. The coefficient of interest, $\beta_1$, captures the differential change in labor-market outcomes for Hong Kong firms following the restriction. To examine heterogeneous effects, we augment the baseline model with interactions between the treatment indicator and firm characteristics, including firm size, workforce seniority, and dependence on external platforms.

Although our setting includes three policy changes: introduction, restriction, and restoration, our baseline analysis focuses on the October 2024 restriction, which directly addresses this study's central research question. We additionally examine the introduction and restoration events as validation tests. We find that labor market outcomes improve following the introduction, deteriorate after the restriction, and recover after the restoration, providing a consistent pattern that strengthens our causal interpretation, as it is difficult for alternative explanations to generate effects that align in both timing and direction with multiple policy changes over the period. We discuss these analyses in Section 5.

A potential concern regarding our DiD design is that treatment is assigned at the country level, while outcomes are observed at the firm-month level. Although the firm-level panel provides substantial within-country variation, the baseline specification effectively relies on a comparison between one treated market (Hong Kong) and one control market (Singapore). Consequently, estimated effects could be influenced by idiosyncratic developments in the control market, and conventional standard errors may understate uncertainty when treatment variation originates from a small number of higher-level clusters.

To address this concern, we conduct a series of robustness analyses. First, we expand the control group to include additional comparable Asian labor markets, including South Korea, Taiwan, and Japan. Second, we estimate standard errors using increasingly conservative clustering schemes, including country-month and country-industry-month clustering. Third, we implement randomization (permutation) inference, repeatedly reassigning treatment status across countries to construct empirical null distributions for our estimates. Across these alternative specifications, the results for both matching quality and matching efficiency remain highly consistent, reducing concerns that our findings are driven by the characteristics of any single control market. The results are listed in Appendix A1.

A second challenge concerns the causal chain linking LinkedIn's policy change to labor-market outcomes. Establishing this mechanism requires demonstrating that: (1) a substantial share of firms rely on LinkedIn as a meaningful recruitment channel; (2) LinkedIn's AI training data access plausibly affects the effectiveness of its AI-enabled matching system; (3) greater reliance on external platform recruitment is more significantly impacted by LinkedIn policy changes, and (4) the observed labor-market effects cannot be explained by concurrent changes in local economic conditions, user behavior, or other market-wide shocks.

To address these concerns, we conduct a series of validations and mechanism analyses summarized in Figure 3. These analyses examine firms' dependence on LinkedIn-based recruiting, the effects of restricting user data for AI training on platform matching algorithm degradation, and a range of alternative explanations, such as time-varying market conditions and user behaviors. We discuss these tests in detail in Section 5.

Figure 3. Causal Chain Diagram: Mechanism Analysis

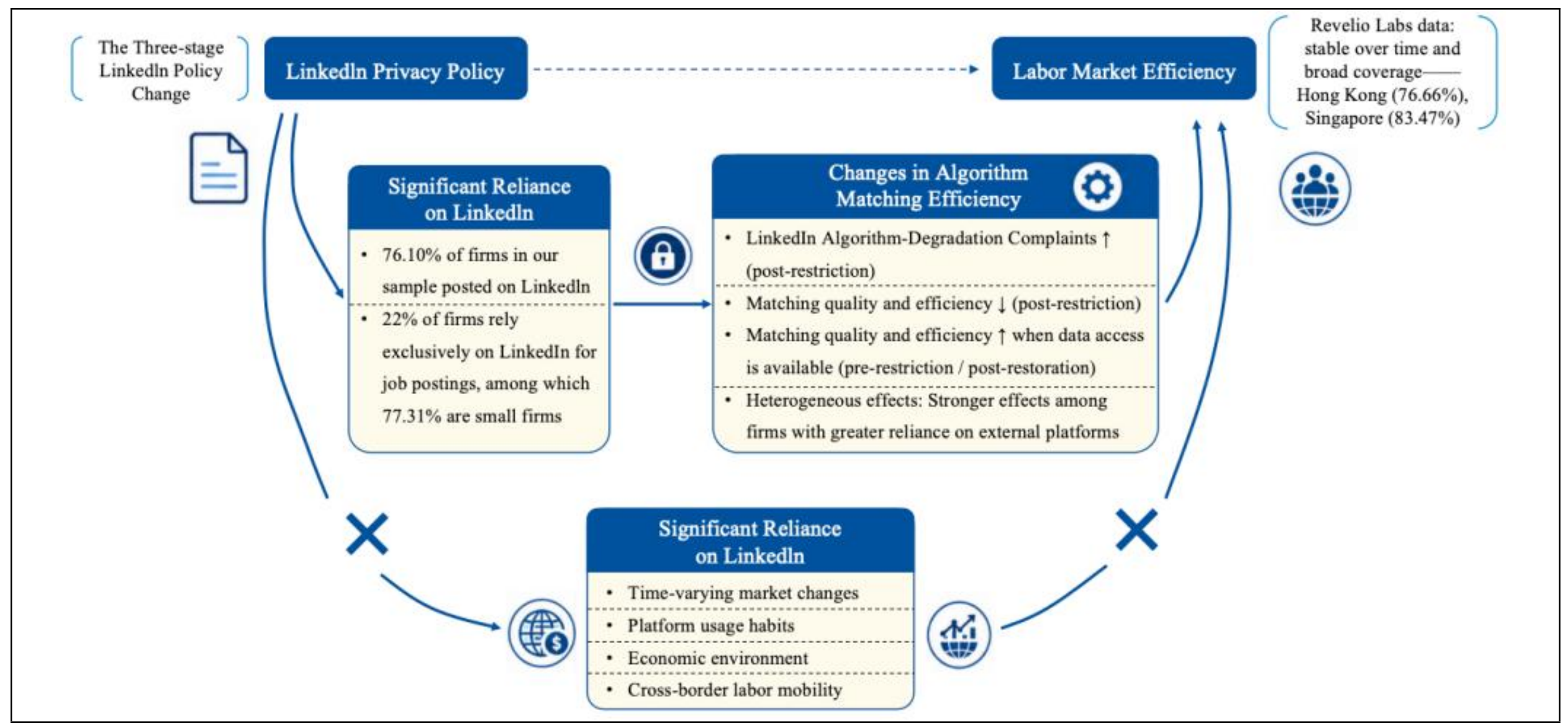

## 4. MAIN RESULTS: MATCHING QUALITY AND EFFICIENCY

We first present the DiD regression results using labor-market matching outcomes as dependent variables. Table 2 reports our baseline DiD estimates for the two matching quality outcomes, Flux Rate and Average Duration, using the specification of equation (1).

Table 2. Main Results — Matching Quality

| Variables | Flux Rate | Average Duration |
|---|---|---|
| Treat × Post | 0.0020*** (0.0004) | -3.7749* (2.0884) |
| Observations | 280,277 | 314,304 |
| R-squared | 0.2142 | 0.9915 |

*Notes: Firm-level clustered standard errors are in parentheses; ***p < 0.01; **p < 0.05; *p < 0.1. Firm- and month-fixed effects are included. The high $R^2$ for Average Duration reflects high-dimension firm fixed effects absorbing persistent cross-firm heterogeneity (Lemmon et al., 2008).*

Column (1) in Table 2 shows that, following the October 2024 restriction, the Flux Rate (the sum of inflows and outflows from outside organizations, scaled by the firm's total number of employees) of companies in Hong Kong rose significantly in comparison to those in Singapore (Treat × Post = 0.0020, $p < 0.01$). Given a sample mean of 0.0136, this estimate corresponds to an increase of approximately 15%, indicating substantially greater workforce turnover following the restriction. Column (2) shows that the Average Duration, which measures the average realized tenure of employees associated with a firm, also declined after the restriction policy (Treat × Post = −3.7749, $p < 0.1$). Because successful employer–employee matches tend to persist longer, while mismatches are more likely to terminate early, the observed decline in employment duration is consistent with a reduction in matching quality. Figure 4 presents the event-study estimates for both outcome variables. The coefficients are close to zero and statistically insignificant throughout the pre-treatment period, indicating no evidence of differential pre-trends between Hong Kong and Singapore. In contrast, the estimated effects emerge only after the October 2024 restriction (event period 0) and persist thereafter, further supporting the causal interpretation of our baseline DiD results.

Figure 4. Event Study Analysis of Matching Quality

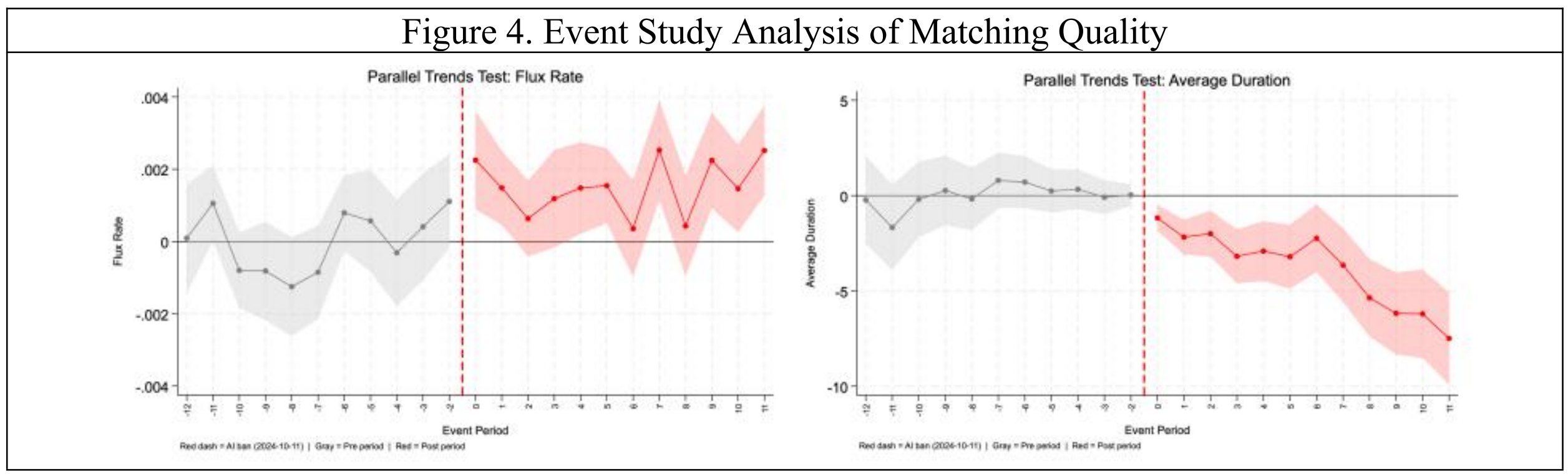

Next, we present the DiD results for matching efficiency in Table 3. Column (1) shows that, following the policy intervention, Active Posting Period increased significantly (Treat × Post = 3.4123, $p < 0.05$), indicating that the time a vacancy remains actively advertised on the platform increased significantly. Relative to the sample mean of 31.04 days, this estimate corresponds to an increase of approximately 3.4 days, implying that vacancies remained advertised for about 11% longer after the restriction. At the same time, Column (2) shows that the Matched Job Postings Rate declined significantly (−0.0141, $p < 0.01$), and Column (3) shows that the Half-Year Matched Job Postings Rate also declined significantly (−0.0075, $p < 0.05$). Relative to their respective sample means of 0.2494 and 0.2069, these estimates translate into reductions of approximately 5.7% and 3.6%, indicating that the posting-to-hire conversion rate declined and that the decline was only partially recovered within a six-month window. Overall, these findings indicate that vacancies remained active on the platform for longer periods and were less likely to be associated with successful hires, consistent with a decline in labor-market matching efficiency following the restriction. Figure 5 presents consistent results in event-study analyses.

Table 3. Direct Evidence on Job Matching Efficiency

| Variables | Active Posting Period | Matched Job Postings Rate | Half-Year Matched Rate |
|---|---|---|---|
| Treat × Post | 3.4123**(1.3335) | -0.0141***(0.0034) | -0.0075**(0.0032) |
| Observations | 27,976 | 365,876 | 365,876 |
| R-squared | 0.4392 | 0.5865 | 0.5212 |

*Notes: Firm-level clustered standard errors are in parentheses; ***p < 0.01; **p < 0.05; *p < 0.1. Firm- and month-fixed effects are included. Active Posting Period is measured from posting date to removal date and is therefore observed only for postings removed before the end of the sample period. The matched-posting outcomes provide complementary evidence that helps interpret longer active posting periods as reflecting reduced matching efficiency.*

Figure 5. Event Study Analysis of Matching Efficiency

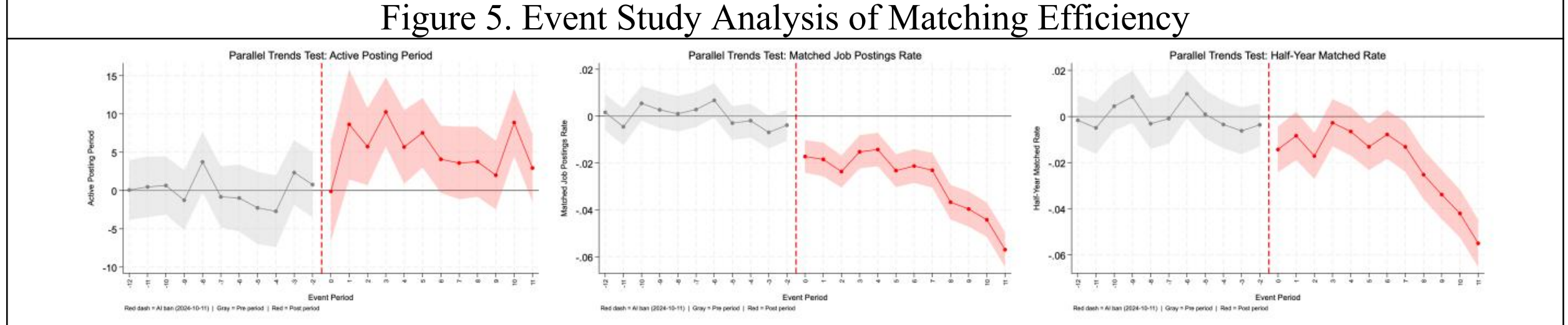

Taken together, the above results indicate that the restriction policy was associated with a decline in both matching quality and matching efficiency: Following the policy change, firms experienced higher workforce turnover, shorter employment durations, longer active posting periods, and lower posting-to-hire conversion rates. While our data do not directly observe LinkedIn's underlying algorithms, the observed pattern is consistent with recent research suggesting that AI models may quickly deteriorate when access to fresh training data is restricted (Shumailov et al., 2024). The findings therefore provide suggestive evidence that limiting the data available for AI training can have downstream consequences for AI-enabled labor-market matching.

## 5. ESTABLISHING THE MECHANISM

The results presented thus far show that labor-market matching quality and efficiency declined in Hong Kong following LinkedIn's restriction on the use of member data for AI training. To strengthen the causal interpretation of these findings, we next examine the mechanism underlying the link between the LinkedIn policy intervention and labor-market outcomes. Specifically, moving from LinkedIn's data policy to firm-level matching outcomes requires establishing four conditions: (1) firms in our setting rely meaningfully on LinkedIn as a recruitment channel; (2) LinkedIn's AI training data access plausibly affects the effectiveness of its AI-enabled matching system; (3) firms that are more exposed to platform-oriented recruitment are more significantly impacted by LinkedIn policy changes, and (4) the observed effects cannot be explained by alternative factors such as changes in local economic conditions, labor-market dynamics, or user behavior.

Because LinkedIn does not disclose internal measures of algorithm performance, we cannot directly observe whether the quality of its matching algorithms deteriorated after the restriction. We therefore assemble multiple complementary pieces of evidence. First, we show that LinkedIn plays a central role in recruitment for a substantial share of firms in our sample, particularly smaller firms. Second, we provide indirect evidence that the policy affected platform matching quality by examining the full policy lifecycle. If LinkedIn's AI-enabled matching depends on access to user data, matching outcomes should improve when AI-training data becomes available, deteriorate when access is restricted, and recover when access is restored. Third, we show that the adverse effects of the data restriction are disproportionately concentrated among smaller firms, less experienced workforces, and firms more dependent on external recruitment platforms, indicating that the policy operates through differential exposure to platform-based AI matching and that the costs fall unevenly on those most reliant on such intermediation. Finally, we conduct a series of tests to rule out alternative explanations, including platform switching, broader labor-market improvements, and other time-varying confounding factors. Taken together, these analyses provide evidence for each link in the proposed causal chain and strengthen the interpretation that the observed deterioration in matching outcomes was driven by LinkedIn's restriction on AI training data rather than by unrelated contemporaneous changes.

### 5.1 Reliance on LinkedIn as a Recruitment Channel

A necessary condition for our proposed mechanism is that firms meaningfully rely on LinkedIn for recruitment. If LinkedIn plays only a marginal role in hiring, changes in its AI-enabled matching capabilities would be unlikely to generate measurable labor-market effects. To assess LinkedIn's importance, we examine recruitment behavior among firms in our sample. Figure 6 reports the distribution of firms' reliance on LinkedIn as a hiring channel. Among firms that posted vacancies during our sample period, over 70% posted at least one vacancy on LinkedIn, indicating that LinkedIn is widely used across both labor markets. More importantly, approximately 22% of sample firms relied exclusively on LinkedIn for recruitment, with no vacancies observed on other job-posting platforms. This exclusive dependence is particularly pronounced among smaller firms, which accounted for 77.31% of all exclusive-LinkedIn firms.

Figure 6. Distribution of Firms' Reliance on LinkedIn as a Recruitment Channel

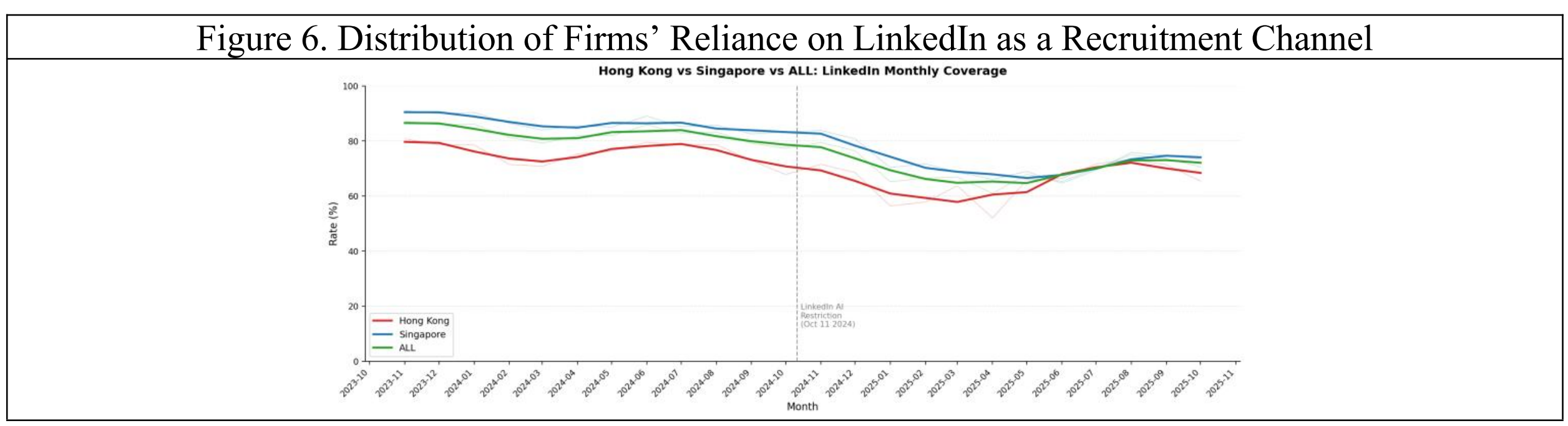

These statistics suggest that LinkedIn is not a peripheral hiring platform but rather a major component of recruitment activity in both markets. For a substantial subset of firms, especially smaller firms, LinkedIn serves as the primary or only observable recruitment channel. Consequently, any deterioration in LinkedIn's matching capability could affect firm-level hiring outcomes and labor-market matching more broadly.

### 5.2 Evidence Linking AI-Training Data Access to Matching Outcomes

Because LinkedIn does not disclose internal measures of algorithm performance, we cannot directly observe whether the restriction on user data access for AI training reduced the quality of its matching algorithms. We therefore rely on multiple complementary pieces of evidence that collectively point to a deterioration in matching performance following the policy intervention.

We first constructed a composite complaint score for algorithm degradation using LinkedIn Apple Store reviews, assigning each review to Hong Kong or Singapore based on IP-derived location. The score measures semantic similarity to pre-defined negative seed sentences about deteriorating recommendation quality, combining standardized character-level TF-IDF, word-level TF-IDF, and keyword density into a z-score. As shown in Figure 7, Hong Kong reviews show a clear increase in the post-restriction period, with the composite score rising above its pre-restriction level, while Singapore remains stable. After data access is restored, Hong Kong's score returns to its pre-restriction level. These results provide suggestive evidence that the restriction was associated with a perceived deterioration in LinkedIn's recommendation quality among Hong Kong users.

Figure 7. LinkedIn Algorithm-Degradation Complaints: Hong Kong vs. Singapore

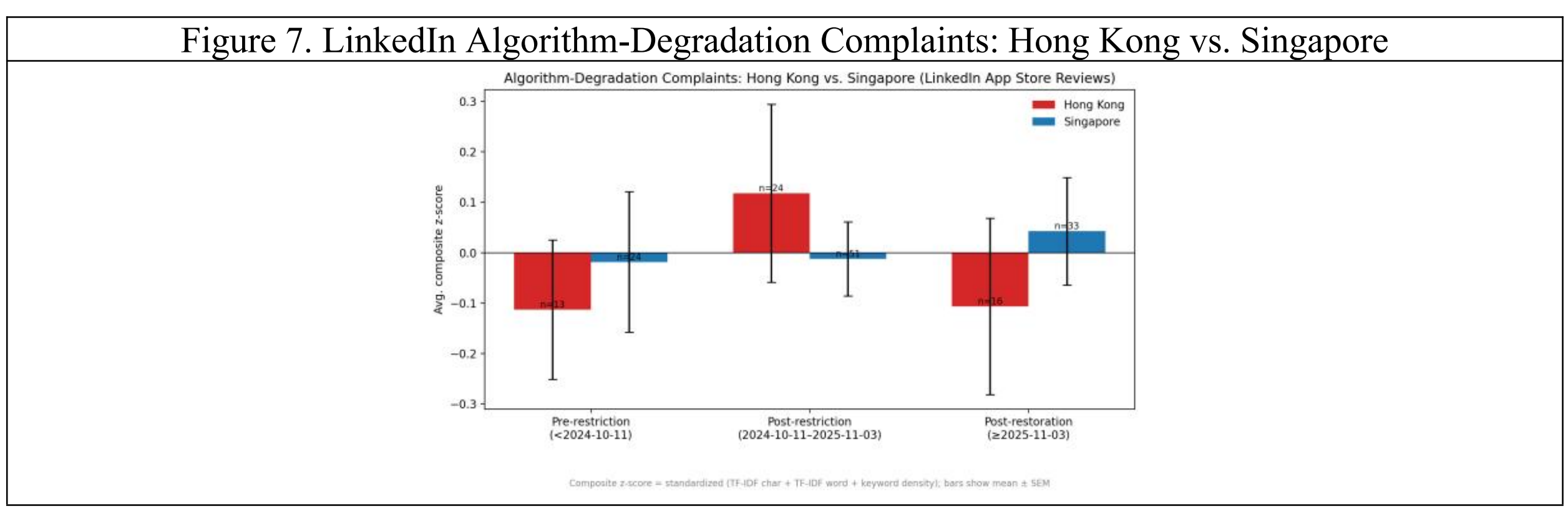

Second, we exploit LinkedIn’s global introduction of user data collection for AI training in August 2024. If access to user data improves AI-enabled matching, labor-market outcomes should improve following the introduction. To test this prediction, we implement a Regression Discontinuity in Time (RDiT) design around the August 2024 policy change. As reported in Table 4, we observe an immediate improvement in matching outcomes in both Hong Kong and Singapore: Flux Rate declines significantly following the introduction, while Average Duration increases significantly. Figure 8 presents the corresponding RDiT plots and shows a clear discontinuity at the policy cutoff. Because this event precedes the Hong Kong restriction, the results suggest that AI-training-data access was already affecting labor-market outcomes before the restriction was imposed. Importantly, the direction of these effects mirrors the pattern observed in our main analyses: matching outcomes improve when AI-training data becomes available and deteriorate when access is restricted. The similarity in magnitude between Hong Kong and Singapore further supports the comparability of the two markets.

Table 4. Improved Labor Market Outcomes at AI Introduction

| Variables | Flux Rate | Average Duration |
|---|---|---|
| Hong Kong | -0.0042*** (0.0008) | 1.5207*** (0.5087) |
| Singapore | -0.0041*** (0.0009) | 1.7644** (0.7576) |

*Notes: Firm-level clustered standard errors are in parentheses; ***p < 0.01; **p < 0.05; *p < 0.1. Firm-fixed effects are included.*

Figure 8. Regression Discontinuity Analysis at the Time of LinkedIn AI Introduction
(Left: Flux Rate; Right: Average Duration)

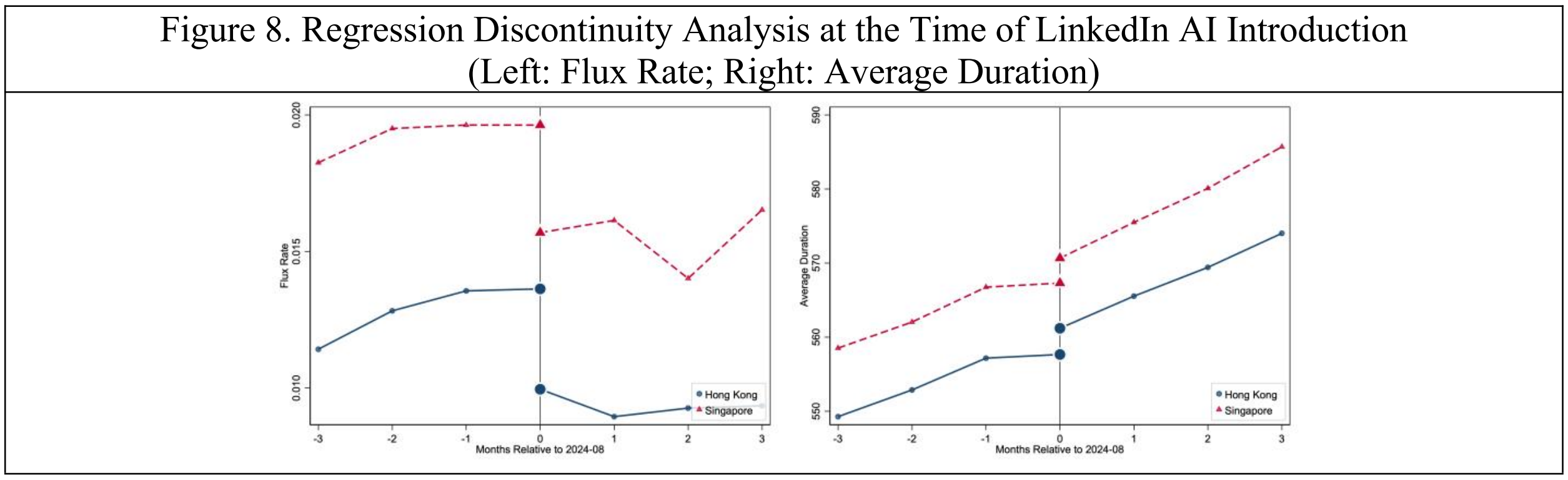

Third, we conduct an external validation using a completely independent institutional setting. Following LinkedIn’s August 2024 introduction of its AI policy, the platform retained access to user data for AI training in the United States and the United Kingdom (Treatment), while such access remained unavailable in the European Union due to GDPR-related restrictions (Control). We therefore compare labor-market outcomes between these two groups. As reported in Table 5, countries with continued access to AI-training data exhibit significantly lower Flux Rates and significantly higher Average Duration than countries without such access. The direction of these effects is the mirror image of our main Hong Kong results. While this setting differs substantially from Hong Kong and Singapore, the same relationship emerges: labor-market matching outcomes improve when AI-training data remain available and deteriorate when such access is restricted. This independent cross-country evidence provides additional support for the role of user data access for AI training in shaping labor-market matching outcomes. These results also enhance the external validity of our findings and suggest that the relationship between user data access for AI training and labor-market matching is not unique to the Hong Kong–Singapore setting.

Table 5. Cross-Country Validation: User Data Access and Labor-Market Outcomes

| Variables | Flux Rate | Average Duration |
|---|---|---|
| Treat × Post | -0.0005** (0.0002) | 4.9290*** (0.4348) |
| Observations | 1,617,062 | 1,845,060 |
| R-squared | 0.2534 | 0.9951 |

*Notes: Firm-level clustered standard errors are in parentheses; ***p < 0.01; **p < 0.05; *p < 0.1. Firm- and month-fixed effects are included. The high $R^2$ for Average Duration reflects the high-dimension firm-fixed effects absorbing persistent cross-firm heterogeneity (Lemmon et al., 2008).*

Fourth, we exploit LinkedIn's restoration of user data access for AI training in Hong Kong as a reverse-treatment test. On November 3, 2025, LinkedIn's updated terms took effect, allowing the platform to resume using Hong Kong users' data for AI training. If the deterioration in matching outcomes following the restriction was driven by reduced access to training data, we should observe a recovery once that access is restored. Figure 9 reports the restoration analysis using the Matched Job Postings Rate, which is the outcome most likely to respond quickly because it captures hiring outcomes at the vacancy stage rather than completed worker transitions.

Figure 9. Event Study Analysis at the Time of LinkedIn AI Training Data Restoration

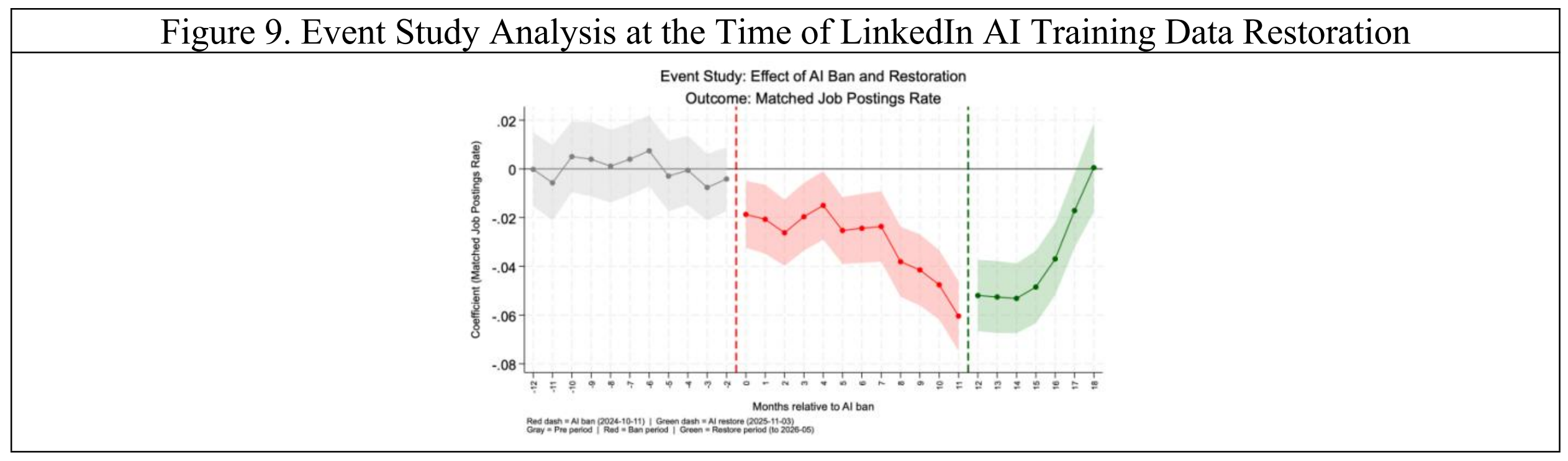

Consistent with our prediction, the Matched Job Postings Rate increases significantly following the restoration. This recovery pattern complements the introduction and restriction results, further supporting the link between AI-training-data access and labor-market matching outcomes. We focus on Matched Job Postings Rate for the restoration analysis because it captures hiring outcomes at the vacancy stage and is therefore most responsive to policy changes. Results for Flux Rate and Average Duration, which we report in Appendix A2, show consistent but more gradual recovery patterns, as expected for employment stock measures.

Taken together, the restoration evidence complements the introduction and restriction results. Matching outcomes improve when AI-training data becomes available, deteriorate when access is restricted, and recover when access is restored. This introduction–restriction–restoration pattern is difficult to reconcile with alternative explanations such as persistent shifts in user preferences, employer migration away from LinkedIn, or other time-varying confounding factors that do not closely track LinkedIn’s AI-training-data policy. As such, the restoration test provides particularly strong evidence linking AI-training-data access to labor-market matching outcomes.

### 5.3 Differential Exposure to Platform-Based Matching

If LinkedIn’s AI-enabled matching system is indeed the mechanism driving our results, firms that rely more heavily on platform-mediated recruitment should be more exposed to the policy shock. We test this prediction using two proxies for platform dependence: firm size and workforce seniority.

### 5.3.1 Firm Size

Table 6 reports the moderating role of firm size. Consistent with the baseline results, the restriction reduces Average Duration and increases Flux Rate. More importantly, the interaction term Treat × Post × Size is positive and significant for Average Duration and negative and significant for Flux Rate. Because firm size is standardized, these estimates indicate that smaller firms experience larger declines in employment duration and greater increases in worker turnover following the restriction.

Table 6. Heterogeneity by Firm Size

| Moderators | Variables | Flux Rate | Average Duration |
|---|---|---|---|
| Firm Size | Treat × Post | 0.0020***(0.0004) | -3.6030***(0.5799) |
| | Treat × Post × Size | -0.0002*(0.0001) | 8.4838***(0.7079) |
| | Observations | 280,277 | 314,304 |
| | R-squared | 0.2142 | 0.9915 |

*Notes: Firm- and month-fixed effects are included; ***p < 0.01; **p < 0.05; *p < 0.1. The high $R^2$ for Average Duration reflects high-dimensional firm fixed effects absorbing persistent cross-firm heterogeneity (Lemmon et al., 2008).*

This pattern is consistent with our proposed mechanism. Smaller firms are generally more reliant on platform-based recruitment, whereas larger firms possess more diversified hiring channels and are therefore less exposed to disruptions in AI-enabled matching. The stronger effects among smaller firms provide additional evidence that the policy operates through platform-mediated recruitment.

### 5.3.2 Workforce Seniority

Table 7 reports the moderating role of workforce seniority. Consistent with the baseline results, the restriction reduces Average Duration and increases Flux Rate. More importantly, the interaction term Treat × Post × Seniority is positive and significant for Average Duration and negative and significant for Flux Rate. Because Seniority is standardized, these estimates indicate that firm-job cells with less experienced workforces experience larger declines in employment duration and greater increases in worker turnover following the restriction.

Table 7. Heterogeneity by Organizational Maturity (Firm-Job-Month Average Seniority)

| Moderators | Variables | Flux Rate | Average Duration |
|---|---|---|---|
| Organizational Maturity | Treat × Post | 0.0013***(0.0003) | -9.6982***(0.4842) |
| | Treat × Post × Seniority | -0.0006**(0.0002) | 7.7602***(0.3853) |
| | Observations | 1,227,216 | 1,227,216 |
| | R-squared | 0.2599 | 0.9892 |

Notes: Firm-job-month level estimation; firm- and month-fixed effects are included; ***$p < 0.01$; **$p < 0.05$; *$p < 0.1$. The estimation is performed at the firm-job-month level to align with the Seniority measure; this finer granularity naturally expands the sample size relative to the firm-month analysis by disaggregating observations across job categories within each firm-month.

This pattern is also consistent with our proposed mechanism. More experienced employees are more likely to rely on professional networks, reputation, and direct recruitment channels, making them less dependent on platform-mediated matching. By contrast, junior employees depend more heavily on online job platforms and algorithmic recommendations. The stronger effects among lower-seniority workforces therefore provide additional evidence that the policy operates through AI-enabled platform matching.

### 5.3.3 Dependence on External Recruitment Platforms

Table 8 reports the moderating role of platform dependence. Consistent with previous mechanism tests, the restriction policy reduces Average Duration and increases Flux Rate more significantly for companies that rely more on external platforms for recruitment.

Table 8. Heterogeneity by Dependence on External Recruitment Platform

| Moderators | Variables | Flux Rate | Average Duration |
|---|---|---|---|
| Dependence on External Platforms | Treat×Post | -0.000 (0.001) | -11.616*** (1.004) |
| | Treat×Post×Dependence | 0.001** (0.000) | -1.553*** (0.524) |
| | Observations | 47,557 | 47,736 |
| | R-squared | 0.1896 | 0.9878 |

Notes: Firm- and month-fixed effects are included; ***$p < 0.01$; **$p < 0.05$; *$p < 0.1$. The sample is restricted to firm-month observations with at least one hire from an external recruitment platform in the preceding three months before the restriction date, as Dependence is undefined otherwise. The high $R^2$ for Average Duration reflects firm fixed effects absorbing persistent cross-firm heterogeneity (Lemmon et al., 2008).

Taken together, the firm size, seniority, and platform dependence results suggest that the costs of the data restriction are not evenly distributed. Instead, they fall disproportionately on firms and worker groups that depend most heavily on platform-mediated recruitment, further supporting the proposed mechanism.

### 5.4 Ruling Out Alternative Explanations

While the evidence presented above is consistent with the AI-training-data mechanism, alternative explanations remain possible. In this section, we examine three prominent alternatives: employer migration away from LinkedIn, an interpretation based on increased labor-market dynamism rather than reduced matching quality, and unrelated concurrent shocks.

#### 5.4.1 Platform Switching

A natural concern is that the observed deterioration in matching outcomes reflects employers shifting recruitment activity away from LinkedIn toward alternative hiring platforms, rather than changes in the effectiveness of LinkedIn's matching system. If this explanation were correct, we would expect LinkedIn job-posting activity to decline following the policy intervention. Table 9 reports the results. We find no statistically significant change in the number of LinkedIn job postings after the restriction. The point estimate is close to zero and economically small. These findings suggest that firms continued to use LinkedIn as a recruitment channel at a similar intensity following the policy change, making employer migration to alternative platforms an unlikely explanation for the observed decline in matching outcomes. We also examine total job postings across all platforms. The coefficient is also insignificant, indicating that aggregate hiring activity remained stable during the restriction period.

Table 9. Test for Platform Switching — Posting Volume

| Variables | All Postings Number | LinkedIn Postings Number |
|---|---|---|
| Treat × Post | -0.1510 (0.1544) | -0.0188 (0.0637) |
| Observations | 1,360,200 | 1,360,200 |
| R-squared | 0.3683 | 0.2721 |

*Notes: Firm-level clustered standard errors are in parentheses; ***p < 0.01; **p < 0.05; *p < 0.1. Firm- and month-fixed effects are included.*

#### 5.4.2 Labor-Market Dynamism versus Labor-Market Friction

Another interpretation of our findings is that the increase in worker mobility reflects greater labor-market dynamism rather than deteriorating matching quality. Under this view, higher turnover could arise because workers face better outside opportunities and are able to move more efficiently across employers. To distinguish between these interpretations, we examine wage outcomes. Greater labor-market dynamism is typically associated with improved matching opportunities and stronger outside options, both of which tend to place upward pressure on wages. By contrast, increased matching frictions should weaken wage outcomes.

Table 10 reports the results. Following the restriction, all three wage measures decline significantly. In terms of economic magnitude, these coefficients correspond to declines of approximately 2% relative to the sample mean for wage. The wage decline is modest but consistent in direction across all three measures. Such evidence is difficult to reconcile with a labor-market dynamism explanation because a more dynamic labor market would be expected to generate higher wages alongside increased mobility. Instead, we observe the opposite pattern: worker flux increases, vacancies remain open longer, fill rates decline, employment durations shorten, and wages fall. This consistent pattern across multiple dimensions supports the interpretation that the policy increased labor-market frictions rather than because of improving labor-market dynamism.

Table 10. Wage Changes Following the Data Restriction Policy

| Variables | Avg. Wage of Posts | Avg. Wage by Job-Senior Level | Firm Sum Wage |
|---|---|---|---|
| Treat × Post | -839.22*** (267.77) | -2,426.5*** (518.17) | -67,054.2* (38900.8) |
| Observations | 365,839 | 314,304 | 314,304 |
| R-squared | 0.5407 | 0.9995 | 0.9997 |

*Notes: Firm- and month-fixed effects are included; ***p < 0.01; **p < 0.05; *p < 0.1. The high $R^2$ reflects firm fixed effects absorbing persistent cross-firm heterogeneity (Lemmon et al., 2008).*

### 5.4.3 Other Robustness Checks

In this section, we list the remaining empirical concerns and corresponding mitigations. Table 11 provides a concise summary of the additional robustness checks.

Table 11. Other Empirical Concerns and Robustness Checks

| Empirical Concern | Robustness Check | Location of the Test |
|---|---|---|
| The estimated effects reflect unrelated shocks that happened to coincide with the | Unrelated shocks are unlikely to drive all the Introduction-Restriction-Restoration | Section 5.2 and Appendix A2 |

| LinkedIn policy change | results | |
|---|---|---|
| The results are driven by existing trends in the market | No effects in placebo test which moves the treatment time to 1 year earlier | Appendix A3 |
| The components of the flux rate may lead to opposite/inconsistent directions | Decompose the flux rate into inflow rate and outflow rate and the results remain consistent | Appendix A4 |
| The estimated effects are driven by external market or macroeconomic condition changes | Control time-varying macroeconomic conditions such as GDP, Rental Price, CPI, etc. | Appendix A5 |
| Cross-border labor mobility between Hong Kong and Singapore may violate the SUTVA assumption | Exclude the companies that have ever hired employees from the other country and the results remain consistent | Appendix A6 |
| Cross-country differences may violate the pre-trend assumption | Conduct Synthetic DiD analysis with reasonable donor weights. The results remain consistent | Appendix A7 |

## 6. CONCLUSION AND DISCUSSION

This study examines the labor-market consequences of restricting access to user data for AI training. Leveraging LinkedIn's introduction, restriction, and subsequent restoration of user data access, we provide causal evidence on how changes in data availability affect AI-enabled labor-market matching. Using large-scale employment and job-posting data from Revelio Labs and a DiD design comparing Hong Kong and Singapore, we find that the restriction significantly increased labor-market frictions. Following the policy change, employee tenure declined, worker flux increased, vacancies remained active for longer periods, job-posting match rates fell, and wages decreased. We further show that these effects are strongest among firms and worker groups that rely more heavily on platform-mediated recruitment. Across multiple validation exercises, including LinkedIn's AI introduction, restoration, external cross-country comparisons, and heterogeneity analyses, the evidence consistently points to a link between AI-training-data access and labor-market matching outcomes.

Our analysis has several limitations. First, we do not directly observe LinkedIn's internal matching algorithms or proprietary performance metrics. Consequently, our conclusions rely on observed labor-market outcomes and a set of complementary validation tests rather than direct measures of algorithm quality. Second, although Hong Kong's policy intervention provides unusually clean identification, it represents a single institutional setting. Although our cross-country validation in Section 5.2 mitigates this concern by showing consistent patterns in a different institutional setting, the magnitude of the effects may differ across platforms, countries, industries, and regulatory environments. Third, our study focuses on labor-market outcomes and therefore does not quantify the privacy benefits associated with restricting access to personal data. Future research could examine both sides of this trade-off simultaneously by evaluating how privacy protections and algorithmic performance jointly affect social welfare.

At the same time, our findings should not be interpreted as suggesting that stronger privacy protections are necessarily undesirable. The labor-market costs documented in this study capture the short- to medium-term consequences of restricting access to AI-training data. Over longer horizons, firms, workers, and platform operators may adapt by developing alternative recruitment channels, improving privacy-preserving AI techniques, or adjusting to new regulatory environments. As a result, some of the observed efficiency losses may be mitigated as markets gradually adjust to the new equilibrium.

Our findings contribute to the emerging literature on AI governance, digital platforms, and information privacy. Existing discussions regarding AI-training-data restrictions are often motivated by concerns about privacy protection or by technical evidence on model performance. We complement this literature by providing causal evidence on the real-world economic consequences of changing AI-training-data availability. In particular, we show that the effects extend beyond platform-level performance metrics and propagate into broader labor-market outcomes, including worker mobility, vacancy filling, employment duration, and wages. More broadly, the study illustrates how data-governance policies can influence economic activity through AI-enabled digital infrastructures.

The findings also carry important managerial and policy implications. For platform operators, the results suggest that access to high-quality and continuously refreshed user data may be an important input into AI-enabled matching systems. Restrictions on such data can generate measurable reductions in matching efficiency, particularly among users and firms that rely heavily on platform-mediated interactions. For firms, especially smaller organizations with limited recruiting resources, dependence on a single digital platform may increase exposure to changes in platform governance and AI policies. Finally, for policymakers, the results highlight that privacy regulation may have economic consequences that extend beyond compliance costs or consumer welfare. While protecting personal data remains an important objective, restrictions on AI-training data may also affect the efficiency of critical economic infrastructures such as labor markets. As governments continue to develop AI governance frameworks, understanding and quantifying these broader economic effects will be increasingly important for designing policies that balance privacy protection with the societal benefits generated by AI-enabled systems.

## REFERENCES

Acquisti, A., Brandimarte, L., & Loewenstein, G. (2015). Privacy and human behavior in the age of information. *Science*, 347(6221), 509-514. https://doi.org/10.1126/science.aaa1465

Acquisti, A., Taylor, C., & Wagman, L. (2016). The economics of privacy. *Journal of Economic Literature*, 54(2), 442-492. https://doi.org/10.1257/jel.54.2.442

Davis, S. J., & Haltiwanger, J. (1999). Gross job flows. *Handbook of Labor Economics*, 3, 2711-2805. https://doi.org/10.1016/S1573-4463(99)30027-4

Davis, S. J., Faberman, R. J., & Haltiwanger, J. C. (2012). Recruiting intensity during and after the Great Recession: National and industry evidence. *American Economic Review*, 102(3), 584-588. https://doi.org/10.1257/aer.102.3.584

Davis, S. J., Faberman, R. J., & Haltiwanger, J. C. (2013). The establishment-level behavior of vacancies and hiring. *Quarterly Journal of Economics*, 128(2), 581–622. https://doi.org/10.1093/qje/qjt002

Faberman, R. J., & Kudlyak, M. (2016). What does online job search tell us about the labor market. *Economic Perspectives*, 40(1), 1-15.

Garg, R., & Telang, R. (2018). To be or not to be linked: Online social networks and job search by unemployed workforce. *Management Science*, 64(8), 3926-3941. https://doi.org/10.1287/mnsc.2017.2784

Ghose, A., Li, B., Macha, M., Sun, C., & Foutz, N. Z. (2026). Privacy choice during crisis. *Management Science*, 72(6), 5068-5082. https://doi.org/10.1287/mnsc.2022.00017

Godinho de Matos, M., & Adjerid, I. (2022). Consumer consent and firm targeting after GDPR: The case of a large telecom provider. *Management Science*, 68(5), 3330-3378. https://doi.org/10.1287/mnsc.2021.4054

Godinho de Matos, M., Ferreira, P., & Belo, R. (2018). Target the ego or target the group: Evidence from a randomized experiment in proactive churn management. *Marketing Science*, 37(5), 793-811.

https://doi.org/10.1287/mksc.2018.1099
Jovanovic, B. (1979). Job matching and the theory of turnover. *Journal of Political Economy*, 87(5, Part 1), 972-990. https://doi.org/10.1086/260808
Kali, R., & Liu, A. Y. (2024). Labor market power and worker turnover. *European Economic Review*, 163, 104701. https://doi.org/10.1016/j.euroecorev.2024.104701
Kokkodis, M., & Ipeirotis, P. G. (2023). The good, the bad, and the unhirable: Recommending job applicants in online labor markets. *Management Science*, 69(11), 6969-6987. https://doi.org/10.1287/mnsc.2023.4690
Kroft, K., & Pope, D. G. (2014). Does online search crowd out traditional search and improve matching efficiency? Evidence from Craigslist. *Journal of Labor Economics*, 32(2), 259-303. https://doi.org/10.1086/673374
Kuhn, P., & Mansour, H. (2014). Is internet job search still ineffective? *The Economic Journal*, 124(581), 1213-1233. https://doi.org/10.1111/ecoj.12119
Kuhn, P., & Skuterud, M. (2004). Internet job search and unemployment durations. *American Economic Review*, 94(1), 218-232. https://doi.org/10.1257/000282804322970779
Lemmon, M. L., Roberts, M. R., & Zender, J. F. (2008). Back to the beginning: persistence and the cross-section of corporate capital structure. *The Journal of Finance*, 63(4), 1575-1608. https://doi.org/10.1111/j.1540-6261.2008.01369.x
Li, Q., Lourie, B., Nekrasov, A., & Shevlin, T. (2022). Employee turnover and firm performance: Large-sample archival evidence. *Management Science*, 68(8), 5667-5683. https://doi.org/10.1287/mnsc.2021.4199
Lu, T., Zhang, Y., & Li, B. (2023). Profit vs. equality? The case of financial risk assessment and a new perspective on alternative data. *MIS Quarterly*, 47(4), 1517-1556. https://doi.org/10.25300/MISQ/2023/17330
Marinescu, I., & Wolthoff, R. (2020). Opening the black box of the matching function: The power of words. *Journal of Labor Economics*, 38(2), 535-568. https://doi.org/10.1086/705903
Pissarides, C. A. (2011). Equilibrium in the labor market with search frictions. *American Economic Review*, 101(4), 1092-1105. https://doi.org/10.1257/aer.101.4.1092
Shumailov, I., Shumaylov, Z., Zhao, Y., Papernot, N., Anderson, R., & Gal, Y. (2024). AI models collapse when trained on recursively generated data. *Nature*, 631(8022), 755-759. https://doi.org/10.1038/s41586-024-07566-y
Sun, T., Yuan, Z., Li, C., Zhang, K., & Xu, J. (2024). The value of personal data in internet commerce: A high-stakes field experiment on data regulation policy. *Management Science*, 70(4), 2645-2660. https://doi.org/10.1287/mnsc.2023.4828
Zhang, X., Ferreira, P., de Matos, M. G., & Belo, R. (2021). Welfare properties of profit maximizing recommender systems: Theory and results from a randomized experiment. *MIS Quarterly*, 45(1), 1-43. https://doi.org/10.25300/MISQ/2021/14971
Zhang, X., Pang, Y., Kang, Y., Chen, W., Fan, L., Jin, H., & Yang, Q. (2025). No free lunch theorem for privacy-preserving llm inference. *Artificial Intelligence*, 341, 104293. https://doi.org/10.1016/j.artint.2025.104293

# APPENDIX

## Appendix A1: Expanded Control Group, Granular Clustering, and Randomized Inference

In addition to Singapore, we incorporate additional control units drawn from the other comparable Asian markets, including "Four Asian Tigers" economies (South Korea and Taiwan) and Japan, which were not subject to LinkedIn's AI data restrictions. We then re-estimate our main specification while clustering standard errors at both the country-month level and the more conservative country-month-industry level. The results remain quantitatively similar to our baseline estimates and continue to be statistically significant.

Table A1. Expand Control group and More Conservative Clustering Schemes

| Variables (Treat × Post) | Flux Rate | Average Duration |
|---|---|---|
| cluster region-month | 0.0002**(0.0001) | -5.0997***(0.7625) |
| cluster region-month-industry | 0.0002***(0.0000) | -5.0997***(0.5994) |
| Observations | 548,971 | 647,720 |
| R-squared | 0.6164 | 0.9811 |

*Notes: ***p < 0.01; **p < 0.05; *p < 0.1. Firm- and Month-fixed effects are included.*

We also implement randomization inference, repeatedly reassigning treatment status at the firm-month-industry level and constructing the empirical distribution of treatment effects under the sharp null hypothesis of no effect. The estimated treatment effects of the data restriction policy on Average Duration and Flux Rate lie in the extreme tails of this empirical distribution, suggesting that the estimated effects are unlikely to be driven by random chance.

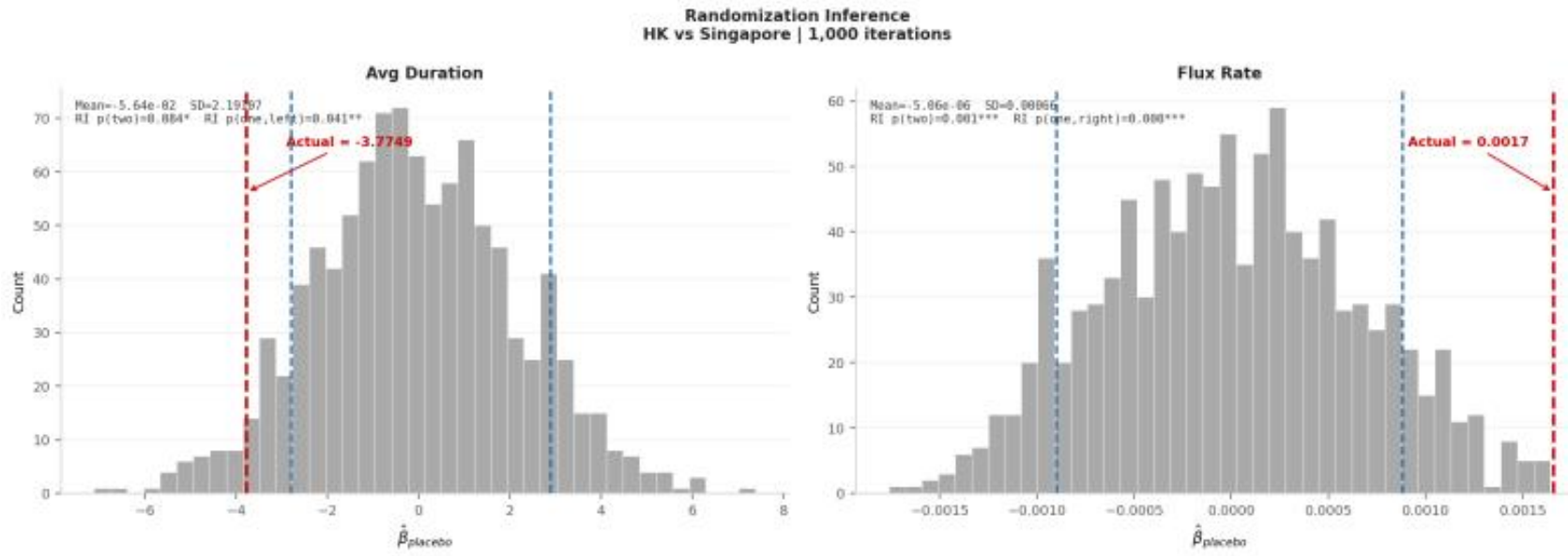

## Appendix A2: Event Study at Policy Restoration

The following figures show the event study of Flux Rate and Average Duration at the time of LinkedIn AI-training data policy restoration (November 2025). The results show patterns consistent with our predictions: labor-market frictions recovered gradually after the policy restoration.

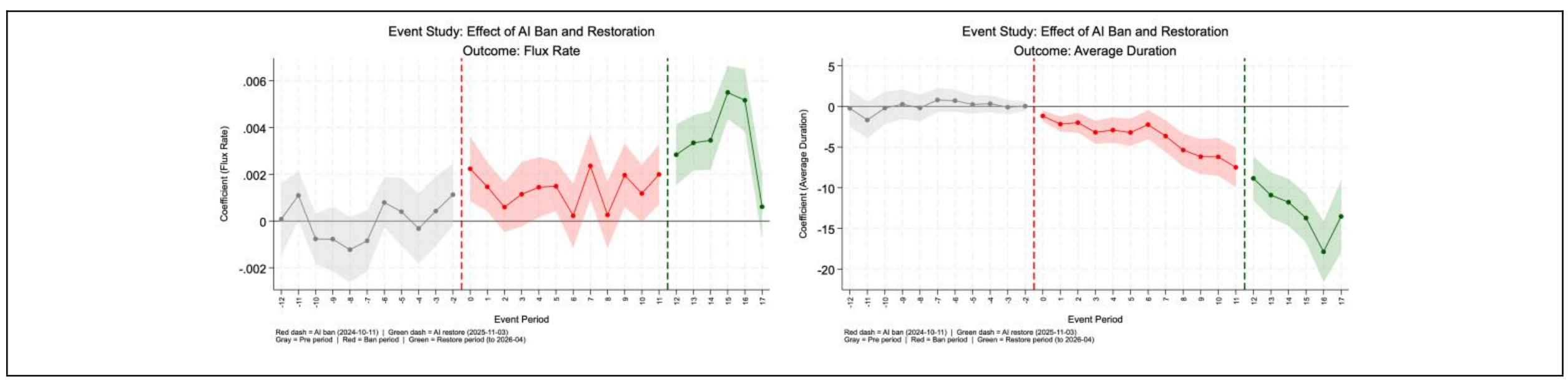

## Appendix A3: Placebo Test by Moving the Treatment Time to 1 Year Earlier

The following table shows that the DiD estimates are no longer statistically significant after moving the treatment time to 1 year earlier (2023/10/11). This result helps rule out potential alternative explanations such as pre-existing labor-market trends.

Table A2. Placebo Test by Moving the Treatment Time to 1 Year Earlier (2023/10/11)

| Variables | Flux Rate | Average Duration |
|---|---|---|
| Treat × Post | -0.0011 (0.0007) | -1.006 (1.540) |
| Observations | 139,576 | 157,152 |
| R-squared | 0.3199 | 0.9938 |

*Notes: Firm-level clustered standard errors are in parentheses; ***p < 0.01; **p < 0.05; *p < 0.1. Firm- and Month-fixed effects are included. The estimation uses a symmetric 6-month window before and after the placebo date (October 11, 2023).*

## Appendix A4: Decomposing Flux Rate into Inflow and Outflow Rate

The following table shows that both inflow rate and outflow rate significantly increased after the restriction policy was implemented, consistent with the baseline regression results.

Table A3. Decomposing Flux Rate into Inflow and Outflow Rate

| Variables | Inflow Rate | Outflow Rate |
|---|---|---|
| Treat × Post | 0.0011*** (0.0003) | 0.0009*** (0.0003) |
| Observations | 280,277 | 280,277 |
| R-squared | 0.1610 | 0.1337 |

*Notes: Firm-level clustered standard errors are in parentheses; ***$p < 0.01$; **$p < 0.05$; *$p < 0.1$. Firm- and Month-fixed effects are included.*

## Appendix A5: Time-Varying Macroeconomic Controls

The following table shows that the DiD estimates remain largely unchanged after adding time-varying macroeconomic control variables, including monthly regional GDP, Rental Price Index, Population, Unemployment Rate, and CPI Growth Rate.

Table A4. Robustness — Time-Varying Macroeconomic Controls

| Variables | Flux Rate | Average Duration |
| --- | --- | --- |
| Treat × Post | 0.0016** (0.0008) | -2.8529** (1.2425) |
| ln_GDP | -0.0209** (0.0083) | -15.6148* (8.2051) |
| ln_rent | 0.0032 (0.0106) | -20.1982*** (5.9073) |
| ln_population | 0.0656 (0.0407) | -52.7037 (47.8142) |
| unemployment_rate | 0.0011 (0.0009) | -4.9618*** (1.7064) |
| CPI_rate | -0.0004 (0.0004) | -0.4724 (0.5258) |
| Observations | 280,277 | 314,304 |
| R-squared | 0.2143 | 0.9915 |

*Notes: Firm-level clustered standard errors are in parentheses; ***$p < 0.01$; **$p < 0.05$; *$p < 0.1$. Firm- and Month-fixed effects are included.*

## Appendix A6: Cross-Border Labor Mobility and SUTVA

The following table shows that the results remain consistent after excluding Hong Kong firms that have ever hired employees from Singapore and Singapore firms that have ever hired employees from Hong Kong.

Table A5. Robustness — Excluding Firms with Cross-Border Hiring Links

| Variables | Flux Rate | Average Duration |
| --- | --- | --- |
| Treat × Post | 0.0020*** (0.0005) | -3.4988*** (2.1196) |
| Observations | 274,469 | 308,496 |
| R-squared | 0.2123 | 0.9915 |

*Notes: Firm-level clustered standard errors are in parentheses; ***$p < 0.01$; **$p < 0.05$; *$p < 0.1$. Firm- and Month-fixed effects are included.*

## Appendix A7: Synthetic DiD

The following figure shows the synthetic DiD results using the placebo-based variance estimator implemented in the synthdid package. The estimated effect on Average Duration is significant at the 1% level ($p < 0.001$); the estimated effect on Flux Rate is significant at the 10% level ($p = 0.0995$). Both point estimates are directionally consistent with the main panel DiD.

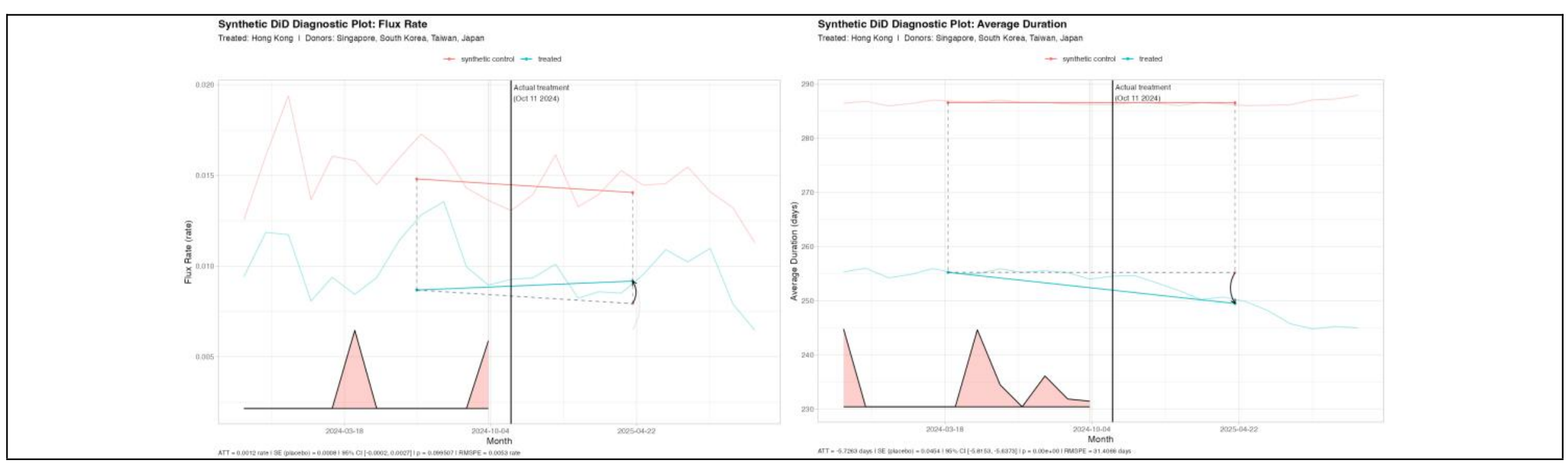

*Notes: The donor pool weights for Flux Rate are Singapore (0.2566), South Korea (0.2278), Taiwan (0.2949), Japan (0.2207). The corresponding weights for Average Duration are Singapore (0.2193), South Korea (0.2467), Taiwan (0.2653), Japan (0.2686).*